\documentclass[english]{revtex4}
\usepackage[T1]{fontenc}
\usepackage[latin1]{inputenc}

\makeatletter

\usepackage{babel}
\makeatother
\begin{document}

\title{Thermodynamics of Charged BTZ Black Holes and Effective String Theory}

\author{Eduard Alexis Larrañaga Rubio}

\email{eduardalexis@gmail.com}

\affiliation{National University of Colombia}

\affiliation{National Astronomical Observatory (OAN)}

\begin{abstract}
In this paper we study the first law of thermodynamics for the (2+1)
dimensional charged BTZ black hole considering a pair of thermodinamical
systems constructed with the two horizons of this solution. We show
that these two systems are similar to the right and left movers of
string theory and that the temperature associated with the black hole
is the harmonic mean of the temperatures associated with these two
systems. 
\end{abstract}
\maketitle

\section{Introduction}

Bekenstein and Hawking showed that black holes have non-zero entropy
and that they emit a thermal radiation that is proportional to its
surface gravity at the horizon. When the black hole has other properties
as angular momentum $\mathbf{J}$ and electric charge $Q$, these
quantities are related with the mass through the identity

\begin{equation}
dM=TdS+\Omega dJ+\Phi dQ,\end{equation}

where $\Omega=\frac{\partial M}{\partial J}$ is the angular velocity
and $\Phi=\frac{\partial M}{\partial Q}$ is the electric potential.
This relation is called \emph{the first law of black hole thermodynamics}\cite{hawking,bekenstein}.
When the black hole has two horizons, it is known that it is possible
to associate a first law with each of them. The outer horizon is related
with the Hawking radiation while the inner horizon is related with
the absortion proccess.\\

In this paper we will apply the method described by Wu \cite{wu2}
to describe the thermodynamics of the charged BTZ black hole in (2+1)
dimensions and relate it with the effective string theory and D-brane
description of black holes. In order to accomplish this, we will define
two thermodynamical systems as the sum and the difference of the two
horizons associated with the rotating BTZ black hole. These systems
resemble the R and L moving modes of string theory and will provide
a way to show how the Hawking temperature $T_{H}$ associated with
the BTZ black hole can be interpreted as the harmonic mean of thetemperature
of the R and L parts, i.e.

\begin{equation}
\frac{2}{T_{H}}=\frac{1}{T_{R}}+\frac{1}{T_{L}}.\end{equation}

\section{The Charged BTZ Black Hole}

The charged BTZ black hole \cite{martinez} is a solution of $\left(2+1\right)$
dimensional gravity with a negative cosmological constant $\Lambda=-\frac{1}{l^{2}}$.
Its line element can be written as

\begin{equation}
ds^{2}=-\Delta dt^{2}+\frac{dr^{2}}{\Delta}+r^{2}d\varphi^{2},\end{equation}

where the lapse function is

\begin{equation}
\Delta=-M+\frac{r^{2}}{l^{2}}-\frac{Q^{2}}{2}\ln\left(\frac{r}{l}\right).\end{equation}

This solution has two horizons given by the condition $\Delta=0$,

\begin{equation}
M=\frac{r_{\pm}^{2}}{l^{2}}-\frac{Q^{2}}{2}\ln\left(\frac{r_{\pm}}{l}\right)\label{eq:radii}\end{equation}

The Bekenstein-Hawking entropy associated with the black hole is twice
the perimeter of the outer horizon,

\begin{equation}
S=4\pi r_{+},\end{equation}
and therefore, the mass can be written as

\begin{equation}
M=\frac{S^{2}}{\left(4\pi\right)^{2}l^{2}}-\frac{Q^{2}}{2}\ln\left(\frac{S}{4\pi l}\right).\label{eq:massformula}\end{equation}

The Bekenstein-Smarr integral mass formula is \cite{larr}

\begin{equation}
M=\frac{1}{2}TS+\frac{1}{2}\Phi Q+\frac{1}{4}Q^{2}\end{equation}

\begin{equation}
M=\kappa\mathcal{P}+\frac{1}{2}\Phi Q+\frac{1}{4}Q^{2}\label{eq:generalIntFirstLaw}\end{equation}

where $\mathcal{P}=\frac{P}{2\pi}$ is the {}``reduced'' perimeter
and $\kappa$ is the surface gravity. Thus, the Hawking-Bekenstein
entropy can be written as

\begin{equation}
S=4\pi\mathcal{P},\end{equation}
and the mass (\ref{eq:massformula}) is given by

\begin{equation}
M=\frac{\mathcal{P}^{2}}{l^{2}}-\frac{Q^{2}}{2}\ln\left(\frac{\mathcal{P}}{l}\right).\end{equation}

Finally, the differential form of the first law for this black hole
takes the form \cite{Akbar}

\begin{equation}
dM=2\kappa d\mathcal{P}+\Phi dQ.\label{eq:generalDiffFirstLaw}\end{equation}

As discussed before \cite{Wu,larr2}, we can associate a thermodynamics
to both outer and inner horizons. The four laws associated with theser
horizons describe the Hawking radiation process as well as the absortion
process. Therefore, the integral and differential mass formulae can
be written for the two horizons,

\begin{eqnarray}
M & = & \frac{\mathcal{P_{\pm}}^{2}}{l^{2}}-\frac{Q^{2}}{2}\ln\left(\frac{\mathcal{P}_{\pm}}{l}\right)\label{eq:massequations}\\
dM & = & 2\kappa_{\pm}d\mathcal{P}_{\pm}+\Phi_{\pm}dQ.\end{eqnarray}
From these relations, is easy to see that the surface gravity and
electrostatic potential at the two horizons are

\begin{eqnarray}
\kappa_{\pm} & = & \frac{1}{2}\left.\frac{\partial M}{\partial\mathcal{P}_{\pm}}\right|_{Q}=\frac{\mathcal{P_{\pm}}}{l^{2}}-\frac{Q^{2}}{4P_{\pm}}=\frac{r_{\pm}}{l^{2}}-\frac{Q^{2}}{4r_{\pm}}\\
\Phi_{\pm} & = & \left.\frac{\partial M}{\partial Q}\right|_{\mathcal{P}_{\pm}}=-Q\ln\left(\frac{\mathcal{P}_{\pm}}{l}\right)=-Q\ln\left(\frac{r_{\pm}}{l}\right),\end{eqnarray}

while the entropy and temperature associated with each horizon are

\begin{eqnarray}
S_{\pm} & = & 4\pi\mathcal{P}_{\pm}\\
T_{\pm} & = & \frac{\kappa_{\pm}}{2\pi}.\end{eqnarray}

From equation (\ref{eq:massequations}) we can obtain to important
relations,

\begin{eqnarray}
M & = & \frac{\mathcal{P}_{+}^{2}+\mathcal{P}_{-}^{2}}{2l^{2}}-\frac{Q^{2}}{4}\ln\left(\frac{\mathcal{P}_{+}\mathcal{P}_{-}}{l^{2}}\right)\label{eq:sum}\\
\frac{\mathcal{P}_{+}^{2}-\mathcal{P}_{-}^{2}}{l^{2}} & = & \frac{Q^{2}}{2}\ln\left(\frac{\mathcal{P}_{+}}{\mathcal{P}_{-}}\right)\label{eq:diff}\end{eqnarray}

Now, using the inner and outer horizons we will define two independient
thermodynamical systems. Following Wu \cite{wu2}, the R-system will
have a reduced perimeter corrspondient to the sum of the inner and
outer perimeters while L-system corresponds to the difference of these
perimeters,

\begin{eqnarray}
\mathcal{P}_{R} & = & \mathcal{P}_{+}+\mathcal{P}_{-}\\
\mathcal{P}_{L} & = & \mathcal{P}_{+}-\mathcal{P}_{-}.\end{eqnarray}

It is important to note that each of these systems carry two hairs
$\left(M,Q\right)$, but we will show that the electric potential
is different for each of them. However, to begin, we will obtain the
thermodynamical relations for these systems and then, we will relate
them with the thermodynamics of the charged BTZ black hole and its
Hawking temperature.

\section{R-System Thermodynamics}

First, we will focus in the R-system. The surface gravity for this
system is

\begin{equation}
\kappa_{R}=\frac{1}{2}\left(\frac{\partial M}{\partial\mathcal{P}_{R}}\right)_{Q}=\frac{1}{2}\left[\left(\frac{\partial M}{\partial\mathcal{P}_{+}}\right)_{Q}\left(\frac{\partial\mathcal{P}_{+}}{\partial\mathcal{P}_{R}}\right)_{Q}+\left(\frac{\partial M}{\partial\mathcal{P}_{-}}\right)_{Q}\left(\frac{\partial\mathcal{P}_{-}}{\partial\mathcal{P}_{R}}\right)_{Q}\right]\label{eq:krfirst}\end{equation}

Since $\mathcal{P}_{R}=\mathcal{P}_{+}+\mathcal{P}_{-}$ , we get

\begin{equation}
\frac{\partial\mathcal{P}_{+}}{\partial\mathcal{P}_{R}}+\frac{\partial\mathcal{P}_{-}}{\partial\mathcal{P}_{R}}=1,\label{eq:aux1}\end{equation}
and equation (\ref{eq:diff}) gives

\begin{equation}
\kappa_{+}\frac{\partial\mathcal{P}_{+}}{\partial\mathcal{P}_{R}}=\kappa_{-}\frac{\partial\mathcal{P}_{-}}{\partial\mathcal{P}_{R}}.\label{eq:aux2}\end{equation}
Hence, using equations (\ref{eq:aux1}) and (\ref{eq:aux2}) we can
write

\begin{eqnarray}
\frac{\partial\mathcal{P}_{+}}{\partial\mathcal{P}_{R}} & = & \frac{\kappa_{-}}{\kappa_{+}+\kappa_{-}}\label{eq:aux3}\\
\frac{\partial\mathcal{P}_{-}}{\partial\mathcal{P}_{R}} & = & \frac{\kappa_{+}}{\kappa_{+}+\kappa_{-}}.\label{eq:aux4}\end{eqnarray}

On the other hand, using equation (\ref{eq:sum}) we obtain

\begin{eqnarray}
\left(\frac{\partial M}{\partial\mathcal{P}_{+}}\right)_{Q} & = & \frac{\mathcal{P}_{+}}{l^{2}}-\frac{Q^{2}}{4\mathcal{P}_{+}}=\kappa_{+}\label{eq:aux5}\\
\left(\frac{\partial M}{\partial\mathcal{P}_{-}}\right)_{Q} & = & \frac{\mathcal{P}_{-}}{l^{2}}-\frac{Q^{2}}{4\mathcal{P}_{-}}=\kappa_{-}.\label{eq:aux6}\end{eqnarray}

Therefore, putting equations (\ref{eq:aux3}),(\ref{eq:aux4}),(\ref{eq:aux5})
and (\ref{eq:aux6}) into (\ref{eq:krfirst}) we obtain

\begin{equation}
\kappa_{R}=\frac{\kappa_{+}\kappa_{-}}{\kappa_{+}+\kappa_{-}}=\frac{\left(\frac{\mathcal{P}_{+}}{l^{2}}-\frac{Q^{2}}{4\mathcal{P}_{+}}\right)\left(\frac{\mathcal{P}_{-}}{l^{2}}-\frac{Q^{2}}{4\mathcal{P}_{-}}\right)}{\frac{\mathcal{P}_{+}+\mathcal{P}_{-}}{l^{2}}-\frac{Q^{2}}{4}\left(\frac{1}{\mathcal{P}_{+}}+\frac{1}{\mathcal{P}_{-}}\right)},\end{equation}
or

\begin{equation}
\frac{1}{\kappa_{R}}=\frac{1}{\kappa_{+}}+\frac{1}{\kappa_{-}}.\label{eq:gravityR}\end{equation}
This equation shows that the temperature for the R-system satisfies

\begin{equation}
\frac{1}{T_{R}}=\frac{1}{T_{+}}+\frac{1}{T_{-}},\label{eq:temperatureR}\end{equation}
while the entropy can be written as

\begin{eqnarray}
S_{R} & = & 4\pi\mathcal{P}_{R}=4\pi\left(\mathcal{P}_{+}+\mathcal{P}_{-}\right)=S_{+}+S_{-}\label{eq:entropyR}\end{eqnarray}

On the other side, the electric potential is obteined by the expression

\begin{eqnarray}
\Phi_{R} & = & \left(\frac{\partial M}{\partial Q}\right)_{\mathcal{P}_{R}}=\left(\frac{\partial M}{\partial\mathcal{P}_{+}}\right)_{\mathcal{P}_{R},Q}\left(\frac{\partial\mathcal{P}_{+}}{\partial Q}\right)_{\mathcal{P}_{R}}+\left(\frac{\partial M}{\partial\mathcal{P}_{-}}\right)_{\mathcal{P}_{R},Q}\left(\frac{\partial\mathcal{P}_{-}}{\partial Q}\right)_{\mathcal{P}_{R}}+\left(\frac{\partial M}{\partial Q}\right)_{\mathcal{P}_{+},\mathcal{P}_{-},Q},\label{eq:firfirst}\end{eqnarray}

Since $\mathcal{P}_{R}=\mathcal{P}_{+}+\mathcal{P}_{-}$ , we have

\begin{equation}
\left(\frac{\partial\mathcal{P}_{+}}{\partial Q}\right)_{\mathcal{P}_{R}}+\left(\frac{\partial\mathcal{P}_{-}}{\partial Q}\right)_{\mathcal{P}_{R}}=0\label{eq:aux7}\end{equation}
and eqution (\ref{eq:diff}) gives

\begin{equation}
\left(\frac{\partial\mathcal{P}_{+}}{\partial Q}\right)_{\mathcal{P}_{R}}=\frac{1}{2}\frac{\left(\Phi_{+}-\Phi_{-}\right)}{\kappa_{+}+\kappa_{-}}\label{eq:aux8}\end{equation}

On the other hand, using equation (\ref{eq:sum}) we obtain

\begin{eqnarray}
\left(\frac{\partial M}{\partial\mathcal{P}_{+}}\right)_{\mathcal{P}_{R},Q} & = & \frac{\mathcal{P}_{+}}{l^{2}}-\frac{Q^{2}}{4\mathcal{P}_{+}}=\kappa_{+}\label{eq:aux9}\\
\left(\frac{\partial M}{\partial\mathcal{P}_{-}}\right)_{\mathcal{P}_{R},Q} & = & \frac{\mathcal{P}_{-}}{l^{2}}-\frac{Q^{2}}{4\mathcal{P}_{-}}=\kappa_{-}\label{eq:aux10}\\
\left(\frac{\partial M}{\partial Q}\right)_{\mathcal{P}_{R},\mathcal{P}_{+},\mathcal{P}_{-}} & = & \frac{1}{2}\left(\Phi_{+}+\Phi_{-}\right)\label{eq:aux11}\end{eqnarray}

Therefore, putting equations (\ref{eq:aux7}),(\ref{eq:aux8}),(\ref{eq:aux9}),(\ref{eq:aux10})
and (\ref{eq:aux11}) into (\ref{eq:firfirst}), we obtain

\begin{equation}
\Phi_{R}=\frac{\left(\Phi_{+}+\Phi_{-}\right)}{2}+\frac{\kappa_{+}-\kappa_{-}}{\kappa_{+}+\kappa_{-}}\frac{\left(\Phi_{+}-\Phi_{-}\right)}{2}.\label{eq:firsecond}\end{equation}
$ $

Finally, the integral and differential mass formulae for the R-system
are

\begin{eqnarray}
M & = & \kappa_{R}\mathcal{P}_{R}+\frac{1}{2}\Phi_{R}Q+\frac{1}{4}Q^{2}\\
dM & = & 2\kappa_{R}d\mathcal{P}_{R}+\Phi_{R}dQ,\end{eqnarray}
that corresponds to what is expected from equations (\ref{eq:generalIntFirstLaw})
and (\ref{eq:generalDiffFirstLaw}).

\section{L-System Thermodynamics}

Now, we will turn our attention to the L-system. The surface gravity
for this system is

\begin{equation}
\kappa_{L}=\frac{1}{2}\left(\frac{\partial M}{\partial\mathcal{P}_{L}}\right)_{Q}=\frac{1}{2}\left[\left(\frac{\partial M}{\partial\mathcal{P}_{+}}\right)_{Q}\left(\frac{\partial\mathcal{P}_{+}}{\partial\mathcal{P}_{L}}\right)_{Q}+\left(\frac{\partial M}{\partial\mathcal{P}_{-}}\right)_{Q}\left(\frac{\partial\mathcal{P}_{-}}{\partial\mathcal{P}_{L}}\right)_{Q}\right]\label{eq:klfirst}\end{equation}

Since $\mathcal{P}_{L}=\mathcal{P}_{+}-\mathcal{P}_{-}$ , we have

\begin{equation}
\frac{\partial\mathcal{P}_{+}}{\partial\mathcal{P}_{L}}-\frac{\partial\mathcal{P}_{-}}{\partial\mathcal{P}_{L}}=1,\label{eq:aux12}\end{equation}
and using the eqution (\ref{eq:diff}) we obtain

\begin{equation}
\kappa_{+}\frac{\partial\mathcal{P}_{+}}{\partial\mathcal{P}_{L}}=\kappa_{-}\frac{\partial\mathcal{P}_{-}}{\partial\mathcal{P}_{L}}.\label{eq:aux13}\end{equation}
Thus, using equations (\ref{eq:aux12}) and (\ref{eq:aux13}) we can
write

\begin{eqnarray}
\frac{\partial\mathcal{P}_{+}}{\partial\mathcal{P}_{L}} & = & \frac{\kappa_{-}}{\kappa_{-}-\kappa_{+}}\label{eq:aux14}\\
\frac{\partial\mathcal{P}_{-}}{\partial\mathcal{P}_{L}} & = & \frac{\kappa_{+}}{\kappa_{-}-\kappa_{+}}.\label{eq:aux15}\end{eqnarray}

On the other hand, equation (\ref{eq:sum}) gives

\begin{eqnarray}
\left(\frac{\partial M}{\partial\mathcal{P}_{+}}\right)_{Q} & = & \frac{\mathcal{P}_{+}}{l^{2}}-\frac{Q^{2}}{4\mathcal{P}_{+}}=\kappa_{+}\label{eq:aux16}\\
\left(\frac{\partial M}{\partial\mathcal{P}_{-}}\right)_{Q} & = & \frac{\mathcal{P}_{-}}{l^{2}}-\frac{Q^{2}}{4\mathcal{P}_{-}}=\kappa_{-}.\label{eq:aux17}\end{eqnarray}

Therfore, putting equations (\ref{eq:aux14}),(\ref{eq:aux15}),(\ref{eq:aux16})
and (\ref{eq:aux17}) into (\ref{eq:klfirst}) we obtain

\begin{equation}
\kappa_{L}=\frac{\kappa_{+}\kappa_{-}}{\kappa_{-}-\kappa_{+}}=\frac{\left(\frac{\mathcal{P}_{+}}{l^{2}}-\frac{Q^{2}}{4\mathcal{P}_{+}}\right)\left(\frac{\mathcal{P}_{-}}{l^{2}}-\frac{Q^{2}}{4\mathcal{P}_{-}}\right)}{\frac{\mathcal{P}_{-}-\mathcal{P}_{+}}{l^{2}}-\frac{Q^{2}}{4}\left(\frac{1}{\mathcal{P}_{-}}-\frac{1}{\mathcal{P}_{+}}\right)},\end{equation}
or

\begin{equation}
\frac{1}{\kappa_{L}}=\frac{1}{\kappa_{+}}-\frac{1}{\kappa_{-}}.\label{eq:gravityL}\end{equation}
This equation shows that the temperature for the L-system satisfies

\begin{equation}
\frac{1}{T_{L}}=\frac{1}{T_{+}}-\frac{1}{T_{-}},\label{eq:temperatureL}\end{equation}
and the entropy is

\begin{eqnarray}
S_{L} & = & 4\pi\mathcal{P}_{L}=4\pi\left(\mathcal{P}_{+}-\mathcal{P}_{-}\right)=S_{+}-S_{-}.\label{eq:entropyL}\end{eqnarray}

On the other side, the electric potential is give by

\begin{eqnarray}
\Phi_{L} & = & \left(\frac{\partial M}{\partial Q}\right)_{\mathcal{P}_{L}}=\left(\frac{\partial M}{\partial\mathcal{P}_{+}}\right)_{\mathcal{P}_{L},Q}\left(\frac{\partial\mathcal{P}_{+}}{\partial Q}\right)_{\mathcal{P}_{L}}+\left(\frac{\partial M}{\partial\mathcal{P}_{-}}\right)_{\mathcal{P}_{L},Q}\left(\frac{\partial\mathcal{P}_{-}}{\partial Q}\right)_{\mathcal{P}_{L}}+\left(\frac{\partial M}{\partial Q}\right)_{\mathcal{P}_{+},\mathcal{P}_{-},Q},\label{eq:filfirst}\end{eqnarray}

Since $\mathcal{P}_{L}=\mathcal{P}_{+}-\mathcal{P}_{-}$ , we have

\begin{equation}
\left(\frac{\partial\mathcal{P}_{+}}{\partial Q}\right)_{\mathcal{P}_{L}}+\left(\frac{\partial\mathcal{P}_{-}}{\partial Q}\right)_{\mathcal{P}_{L}}=0\label{eq:aux18}\end{equation}
and using the eqution (\ref{eq:diff}) is easily to obtain

\begin{equation}
\left(\frac{\partial\mathcal{P}_{+}}{\partial Q}\right)_{\mathcal{P}_{L}}=\frac{1}{2}\frac{\left(\Phi_{+}-\Phi_{-}\right)}{\kappa_{+}-\kappa_{-}}\label{eq:aux19}\end{equation}

On the other hand, using equation (\ref{eq:sum}) we get

\begin{eqnarray}
\left(\frac{\partial M}{\partial\mathcal{P}_{+}}\right)_{\mathcal{P}_{L},Q} & = & \frac{\mathcal{P}_{+}}{l^{2}}-\frac{Q^{2}}{4\mathcal{P}_{+}}=\kappa_{+}\label{eq:aux20}\\
\left(\frac{\partial M}{\partial\mathcal{P}_{-}}\right)_{\mathcal{P}_{L},Q} & = & \frac{\mathcal{P}_{-}}{l^{2}}-\frac{Q^{2}}{4\mathcal{P}_{-}}=\kappa_{-}\label{eq:aux21}\\
\left(\frac{\partial M}{\partial Q}\right)_{\mathcal{P}_{L},\mathcal{P}_{+},\mathcal{P}_{-}} & = & \frac{1}{2}\left(\Phi_{+}+\Phi_{-}\right)\label{eq:aux22}\end{eqnarray}

Hence, putting equations (\ref{eq:aux18}),(\ref{eq:aux19}),(\ref{eq:aux20}),(\ref{eq:aux21})
and (\ref{eq:aux22}) into (\ref{eq:filfirst}), we have the electric
potential

\begin{equation}
\Phi_{L}=\frac{\left(\Phi_{+}+\Phi_{-}\right)}{2}+\frac{\kappa_{+}+\kappa_{-}}{\kappa_{+}-\kappa_{-}}\frac{\left(\Phi_{+}-\Phi_{-}\right)}{2}.\label{eq:filsecond}\end{equation}

Finally, the integral and differential mass formulae for the L-system
are

\begin{eqnarray}
M & = & \kappa_{L}\mathcal{P}_{L}+\frac{1}{2}\Phi_{L}Q+\frac{1}{4}Q^{2}\\
dM & = & 2\kappa_{L}d\mathcal{P}_{L}+\Phi_{L}dQ,\end{eqnarray}
that corresponds to equations (\ref{eq:generalIntFirstLaw}) and (\ref{eq:generalDiffFirstLaw}).

\section{Relationship between the R,L-systems and the BTZ thermodynamics}

The thermodynamic laws of the R, L- systems are related with the BTZ
black hole thermodynamics. Equations (\ref{eq:gravityR}) and (\ref{eq:gravityL})
can be resumed into 

\begin{equation}
\frac{1}{\kappa_{R,L}}=\frac{1}{\kappa_{+}}\pm\frac{1}{\kappa_{-}},\end{equation}
that corresponds exactly with the relation found by Wu\cite{wu2}
for Ker-Newman black hole and A. Larranaga\cite{larr2} for the BTZ
black hole. This relation is in direct correspondence to effective
string theory and D-brane physics. 

Since temperature is proportional to surface gravity, we have a similar
expression obtained for equations (\ref{eq:temperatureR}) and (\ref{eq:temperatureL}),

\begin{equation}
\frac{1}{T_{R,L}}=\frac{1}{T_{+}}\pm\frac{1}{T_{-}}.\label{eq:temperatures}\end{equation}

This last relation give us immediately an expression for the Hawking
temperature associated with the BTZ black hole, that corresponds to
the temperature of the outer horizon,

\begin{equation}
T_{H}=T_{+}=\frac{\kappa_{+}}{2\pi},\end{equation}

in terms of the temperatures of the R and L systems. The relation
is 

\begin{equation}
\frac{2}{T_{H}}=\frac{1}{T_{R}}+\frac{1}{T_{L}},\end{equation}

which shows that the Hawking temperature is again the harmonic mean
of the R and L temperatures. \\
Now, lets play some attention to the electric potential. Note that
using equations (\ref{eq:gravityR}) and (\ref{eq:gravityL}) we can
rewrite the R and L electric potentials given by (\ref{eq:firsecond})
and (\ref{eq:filsecond}), as

\begin{equation}
\Phi_{R,L}=\frac{\left(\Phi_{+}+\Phi_{-}\right)}{2}+\frac{\kappa_{R,L}}{\kappa_{L,R}}\frac{\left(\Phi_{+}-\Phi_{-}\right)}{2}\label{eq:ficomplete}\end{equation}

This equation is exactly the same found by Wu \cite{wu2} for the
Kerr-Newman black hole, showing again that the effective string theory
thermodynamics seems to be a universal picture holding also in 2+1
gravity.

\section{conclusion}

In this paper we have shown that the thermodynamics of the (2+1) dimensional
charged BTZ black hole can be constructed from two independient thermodynamical
systems that resemble the right and left modes of string theory. If
one assume that the effective strings have the same mass and electric
charge that the charged BTZ black hole, there is a correspondence
between the R and L modes thermodynamics and the thermodynamics of
the horizons. 

We have show that the Hawking temperature associated with the black
hole is obtained as the harmonic mean of the temperatures associated
with the R and L systems, just as in the case of stringy thermodynamics.
Moreover, equation (\ref{eq:ficomplete}) shows that the electric
potential of the R and L systems si related with the electric potential
of the inner and outer horizons with the same quation obtained by
Wu \cite{wu2} for the Kerr-Newman black hole. \\

All these facts suggest that there is a deep connection between string
theory and D-branes with black holes physics that seems to hold in
many cases, not only in General Relativity but also in 2+1 gravity.
Therefore, it is really interesting to investigate if this relation
can give some clue for the understanding of the origin of black hole
entropy.

\end{document}